\documentclass[twocolumn,showpacs,preprintnumbers,amsmath,amssymb,pra]{revtex4}
\usepackage{color}

\usepackage{graphicx}% Include figure files
\usepackage{dcolumn}% Align table columns on decimal point
\usepackage{bm}% bold math
\usepackage{epstopdf}
\usepackage{bbold}
\usepackage{amsthm}
\usepackage{amsmath}
\usepackage{amssymb}
\newtheorem{theorem}{Theorem}

\newcommand{\iden}{\mathbb{1}}
\newcommand{\incoh}{\mathcal{I}}
\newcommand{\io}{\mathcal{IO}}
\newcommand{\tr}{\mathrm{tr}}
\newcommand{\wit}{\mathcal{W}_{nor}}
\newcommand{\rot}{\mathcal{R}}

\begin{document}

\preprint{}

\title{Quantitative Coherence Witness for Finite Dimensional States}% Force line breaks with \\

\author{Huizhong Ren}
\author{Anni Lin}
\author{Siying He}
\author{Xueyuan Hu}
\email{xyhu@sdu.edu.cn}
\affiliation{School of Information Science and Engineering, Shandong University, Jinan 250100, China}%Lines break automatically or can be forced with \\

%\affiliation{Institute of Physics, Beijing National Laboratory for Condensed Matter Physics, Chinese Academy of Sciences, Beijing 100190, China}
%\affiliation{$^3$School of Physical Sciences, University of Chinese Academy of Sciences, Beijing 100049, China}

\date{\today}% It is always \today, today,

\begin{abstract}
We define the stringent coherence witness as an observable which has zero mean value for all of incoherent states and hence a nonzero mean value indicates the coherence. The existence of such witnesses are proved for any finite-dimension states. Not only is the witness efficient in testing whether the state is coherent, the mean value is also quantitatively related to the amount of coherence contained in the state. For an unknown state, the modulus of the mean value of a normalized witness provides a tight lower bound of $l_1$-norm of coherence in the state. When we have some previous knowledge of the state, the optimal witness is derived such that its measured mean value, called the witnessed coherence, equals to the $l_1$-norm of coherence. One can also fix the witness and implement some incoherent operations on the state before the witness is measured. In this case, the measured mean value cannot reach the witnessed coherence if the initial state and the fixed witness does not match well. Based this result, we design a quantum coherence game. Our results provides a way to directly measure the coherence in arbitrary finite dimension states and an operational interpretation of the $l_1$-norm of coherence.
%Coherence witness provides methods to detect coherence in experimental situations where one could avoid to gather full data to get all information of the state. If one uses a witness experiment to get the expectation value which isn't equal to zero, one can definitely draw the conclusion that the state is coherent. On the other hand, if the expectation value is equal to zero, one can not make sure that whether the state is coherent or not. In this parper, we emphasize to figure out that the witness can always measure the coherence of an unknown state with incoherent unitary operations. What's more, the bigger the expectation value is, the more coherence the state contained. Based on this statement, we design a quantum game which uses quantum coherence as a resource to gain the maximal average payoff.
\end{abstract}

\maketitle

\section{Introduction}
Instead of state tomography, the existence of entanglement can be tested by measuring only one observable, called entanglement witness \cite{Lewenstein2000Characterization}: if the average value is negative, the state must be entangled. Moreover, if the measured average value is well blow zero, one can also infer that the entanglement is very large \cite{Eisert2006Quantitative,Brand2005Quantifying}. Various quantum games are designed based entanglement witness, indicating that all entangled states can be used as a resource in these games \cite{PhysRevLett.108.200401,PhysRevLett.110.060405,PhysRevA.87.032306,PhysRevA.95.052326}. Nevertheless, using entanglement witness to test the entanglement remains challenging both experimentally and theoretically, partly because the optimization problem is hard \cite{Lewenstein2000Optimization,PhysRevLett.118.110502}.

Quantum coherence \cite{arXiv:1609.02439,arXiv:1703.01852}, is a more fundamental property in quantum theory and closely related to the resource theory of quantum entanglement \cite{PhysRevLett.115.020403,Hu16,PhysRevLett.116.160407,PhysRevLett.117.020402}. On the prefixed incoherent basis $\{|j\rangle\}$, the incoherent states \cite{Baumgratz2014Quantifying} are defined as those with diagonal density matrix $\incoh:=\{\rho_I:\rho_I=\sum_jp_j|j\rangle\langle j|\}$, and the incoherent operations \cite{Baumgratz2014Quantifying} are those with incoherent Kraus decompositions $\io:=\{\Lambda^I:\ \Lambda^I(\cdot)=\sum_nK_n(\cdot)K_n^\dagger,\ \mathrm{s.t.}\ K_n\incoh K_n^\dagger\subset\incoh\}$. Different coherence measures has been proposed \cite{Baumgratz2014Quantifying,arXiv:1506.07975,PhysRevA.92.022124,Napoli2016Robustness,PhysRevA.94.060302} and their monotonicity under incoherent operations has was analyzed \cite{PhysRevLett.118.060502,PhysRevA.93.012110,PhysRevA.94.012326}. Among those measures, the $l_1$-norm of coherence has ideal properties, such as strong monotonicity \cite{Baumgratz2014Quantifying} and computational simplicity, but lacks an operational interpretation.

Inspired by the entanglement witness, the coherence witness was proposed and proved to be related to the randomness of coherence and $l_1$-norm of coherence for certain classes of states \cite{Napoli2016Robustness,PhysRevA.93.042107}. Similar to entanglement witness, the coherence witness was defined as an observable whose average value is nonnegative for incoherent states and hence a negative average value indicate the existence of coherence. For each coherent state, the optimal witness that reaches the minimal negative average value is proved to exist. A recent experiment \cite{PhysRevLett.118.020403} measures the average value of the optimal witness, and find that for a class of single-qubit states, its opposite coincides with the robustness of coherence as well as the $l_1$-norm of coherence.

In this paper, we prove the existence of a more stringent coherence witness for any finite-dimension coherent state, and solve several optimization problems. Different from the traditional witnesses, our coherence witness has zero average value for all of the incoherent states, and hence a nonzero average value, no matter positive or negative, infers the coherence. This witness is efficient, in the sense that most of coherent states can be witnessed by only one witness, and the number of unsure states reduces fast as the number of witnesses increases. Another advantage of our witness is that it simplifies the optimization problems. For any given normalized coherence witness $W$ and its measurement mean value $c$, we prove that the coherence $C_{l_1}$ in the measured states is at least $|c|$. When a state $\rho$ is known, we derive the explicit form of the optimal witness (which is a normalized witness with maximum average value in $\rho$), and prove that the mean value of optimal witness is just the $l_1$-norm of coherence in $\rho$. This connection builds an operational interpretation of $C_{l_1}$. Intuitively, the measurement of some witnesses is very difficult due to the experimental constraints, so it is important to study the following problem: if we fix the witness $W$, can we optimize the state using incoherent operations such that the mean value of $W$ can reach that of the optimal witness? We give a positive answer when a qubit state is considered, but a negative one in high-dimension case. Based on this result, we design a quantum game where quantum coherence is a resource but some amount of coherence cannot be activated in this game.

\section{Existence of Coherence Witness}
%From Hahn-Banach Theorem \cite{Encyclopedias2015Encyclopedia}, the coherence witness must exist.

Entanglement witness is proved to exist for any entangled states based on Hahn-Banach Theorem and the convexity of the separable states \cite{Horodecki2000Separability}. Following the similar idea, the coherence witness was recently proposed \cite{Napoli2016Robustness}. For any state $\rho\notin\incoh$, there is a Hermit operator $\tilde W$ such that $\tr(\tilde W\rho_I)\geq0$ for all incoherent state $\rho_I\in\incoh$ but $\tr(\tilde W\rho)<0$. Hence $\tilde W$ is called a coherence witness, namely, detecting a negative mean value of $\tilde W$ reveals that $\rho$ contains the quantum coherence.

When single-qubit states are considered, the Bloch representation provides a geometric picture for the coherence witness. Let $\boldsymbol\sigma:=(\sigma_1,\sigma_2,\sigma_3)$ are Pauli matrices, any single-qubit state $\rho=\frac12(\iden+\boldsymbol b\cdot\boldsymbol \sigma)$ is presented as a three-dimension real vector $\boldsymbol b$ with norm $b\leq1$ and all of the incoherent states are on the $z$-axis. For a Hermit operator $W=w_0\iden+\boldsymbol w\cdot\boldsymbol \sigma$, the states satisfying $\tr(W\rho)\equiv w_0+\boldsymbol w\cdot\boldsymbol b=0$ consist a plane in the Bloch space. When $W$ is a coherence witness, all of the states on the $z$-axis is on one side of the plane, and the coherence of the states on the other side can be witnessed by $\tilde W$. In order to witness more coherent states, we should push the plane $\tr(W\rho)=0$ closer to the $z$-axis. The best we can do is that the $z$-axis is just on the plane. This leads to an interesting result that $\tr(W\rho_I)=0,\ \forall\rho_I\in\incoh$, so $\tr(W\rho)\neq0$ indicates the existence of quantum coherence in $\rho$.

Inspired by the above observation, we propose a more stringent coherence witness and prove its feasibility for any finite-dimension states.
\begin{theorem}\label{th:exist}
For any finite-dimension coherent state $\rho$, there exist a Hermit operator $W$ which satisfies \\
(C1) $\tr(W\rho)\neq0$,\\
(C2) $\tr(W\rho_I)=0,\forall \rho_I\in\incoh$.
\end{theorem}
Here and following, a coherence witness of state $\rho$ means the Hermit operator $W$ satisfying the above two conditions. The origin for the existence of the stringent coherence witness is that, the dimension of the state space $\mathcal{D}(\mathcal H_d)$ is $d^2-1$ but all of the incoherent states live in a $(d-1)$-dimension subspace. A precise proof of theorem \ref{th:exist} goes as follows.
\begin{proof}
Any $d$-dimension state $\rho$ can be written as \cite{1751-8121-41-23-235303}
\begin{equation}
\label{rho_d_gen}\rho=\frac{\iden}{d}+\frac{1}{2}\sum_{j=0}^{d-2}\sum_{k=j+1}^{d-1}{({b_s^{jk}}\sigma_s^{jk}+{b_a^{jk}}\sigma_a^{jk})} +\frac12\sum_{l=1}^{d-1}{b^l\sigma^l},
\end{equation}
Here $b_{s,a}^{jk}=\tr(\rho\sigma_{s,a}^{jk}),\ b^l=\tr(\rho\sigma^{l})$, and
\begin{eqnarray}
\sigma_s^{jk}&=&|j\rangle\langle k|+|k\rangle\langle j|,\\
\sigma_a^{jk}&=&-i|j\rangle\langle k|+i|k\rangle\langle j|,\\
\sigma^l&=&\sqrt{\frac{2}{l(l+1)}}\left(\sum_{j=0}^{l-1}|j\rangle\langle j|-l|l\rangle\langle l|\right),
\end{eqnarray}
are the Gell-Mann matrices \cite{1751-8121-41-23-235303} (or standard $SU(d)$ generators). Apparently, states with $b_{s,a}^{jk}=0\ (\forall j,k)$ consist the set of incoherent states, while one of a non-zero $b_{s,a}^{jk}$ indicates the state $\rho$ is coherent.

Similarly, we write a $d$-dimension Hermit operator as $W=w_0\iden+\sum_{jk}{({w_s^{jk}}\sigma_s^{jk}+{w_a^{jk}}\sigma_a^{jk})}+\sum_{l=1}^{d-1}{w^l\sigma^l}$, where $w_0,\ w_{s,a}^{jk},\ w^l$ are real numbers. When $W$ is a coherence witness, $w_0=w^l=0$ from the condition (C2). Hence the general form of a coherence witness is
\begin{equation}
\label{w_d_gen}W=\sum_{jk}{({w_s^{jk}}\sigma_s^{jk}+{w_a^{jk}}\sigma_a^{jk})}.
\end{equation}
Condition (C1) then becomes
\begin{equation}
\label{eq:c1}\sum_{jk}{{w_s^{jk}}b_s^{jk}+{w_a^{jk}}b_a^{jk}}\neq0,
\end{equation}
for at least one non-zero $b_{s,a}^{jk}$. This can be easily satisfied by choosing a coefficient $w_{s,a}^{jk}\neq0$ for a non-zero $b_{s,a}^{jk}$ and other coefficients in Eq. (\ref{w_d_gen}) to be zero. Thus, the coherence witness $W$ exist for any finite-dimension coherent state $\rho$.
\end{proof}

%\section{The efficiency of witnessing the coherence of an unknown state}
The coherence witness is much more efficient than quantum state tomography in judging whether an \emph{unknown} state is coherent or not. In quantum state tomography, one need to measure the mean values of $d^2$ observables to construct a $d$-dimension density matrix; only after that, can one determine whether a state is coherent. However, when one measures a coherence witness $W_1$, the mean value $\tr(W_1\rho)\neq0$ is enough to witness the coherence. If the measurement result $\tr(W_1\rho)=0$ is obtained, one need to measure another witness $W_2$, and so on.

As an example, we consider the single-qubit case. Most of the coherence states can be witnessed by $W_1=\sigma_x$ except those on the $yoz$ plane, i.e., $\rho_{yz}=\frac12(\iden+b_2\sigma_y+b_3\sigma_z)$. For such states, we choose $W_2=\sigma_y$ and have $\tr(W_2\rho_{yz})=b_2$, which vanishes if and only if $\rho_{yz}$ is incoherent. Hence we need measure at most two witnesses to judge wether an known single-qubit state is coherent or not.

When generalize to the $d$-dimension case, we choose the witnesses as $W_{s,a}^{jk}=\sigma_{s,a}^{jk}$ and measure them in turn. Precisely, we first measure $W_s^{12}$ under the state as in Eq. (\ref{rho_d_gen}) and obtain $\tr(W_s^{12}\rho)=b_s^{12}$. If $b_s^{12}\neq0$, we claim that the state is coherent. The only coherent states that cannot be witnessed by $W_s^{12}$ are those with $b_s^{12}$ strictly equal to zero. The proportion of such states is quite small in the set of coherent states. If the mean value of the first witness $W_s^{12}$ is zero, we need to measure the second witness $W_s^{13}$ and claim the coherence of the state on the result $\tr(W_s^{13}\rho)=b_s^{13}\neq0$. In this second step, most of the rest coherent states are witnessed. Almost all of the coherent states can be witnessed by only a few coherence witnesses, although all of the $d(d-1)$ witnesses have to be measured to determinately judge a state is incoherent.

%If unfortunately the unknown state is the one with $b_a^{d-1,d}\neq0$ but $b_{s,a}^{jk}=0$ otherwise, we have to measure all of the $d(d-1)$ witnesses.

%This is the optimal strategy, because a state is incoherent iff all of the $d(d-1)$ independent coefficients $b_{s,a}^{jk}$ are zero, and in order to determine this,

\section{Quantitative coherence witness}
Now we have shown that for a given coherence witness $W$, the nonzero mean value $\tr(W\rho)$ indicate the existence of coherence. Consequently, we explore the problem whether the amount of $\tr(W\rho)$ is quantitatively related to the coherence measure. Here we employ the $l_1$-norm of coherence as a coherence measure
\begin{equation}
C_{l_1}(\rho)=\sum_{i\neq j}\big|\langle i|\rho|j\rangle\big|.
\end{equation}
Our problem is stated as
\begin{equation}\label{eq:pro}
\begin{array}{ll}
\mathrm{minimize} & C_{l_1}(\rho),\\
\mathrm{subject\ to} & \tr(W\rho)=c,\ \rho\geq0,\ \tr(\rho)=1.
\end{array}
\end{equation}
Because multiplying $W$ by a constant real number will cause a change in the mean value, we need to ``normalize'' the witness in order to get a meaningful relationship between $C_{l_1}(\rho)$ and $c$. A normalized coherence witness should satisfy (N1) $\tr(W\rho)\leq d-1$, and (N2) $\tr(W\rho)= d-1$ only when $\rho$ is a $d$-dimension maximal coherent state. Direct calculation shows that a normalized coherence witness can be written as
\begin{equation}
W=\sum_{jk}(\cos\theta_{jk}\sigma_s^{jk}+\sin\theta_{jk}\sigma_a^{jk}).
\end{equation}
where $\theta_{jk}\in[0,2\pi)$ are called the orientation of $W$. The set of normalized coherence witness is labeled as $\wit$. By writing the state $\rho$ as in Eq. (\ref{rho_d_gen}) and using the labels $\boldsymbol{\omega_{jk}}=(\cos\theta_{jk},\sin\theta_{jk})$ and $\boldsymbol{b_{jk}}=(b_s^{jk},b_a^{jk})$, we have
\begin{equation}
\tr(W\rho)=\sum_{jk}\boldsymbol{\omega_{jk}}\cdot\boldsymbol{b_{jk}},
\end{equation}
and
\begin{equation}
C_{l_1}(\rho)=\sum_{jk}\sqrt{(b_s^{jk})^2+(b_a^{jk})^2}=\sum_{jk}|\boldsymbol{b_{jk}}|.
\end{equation}
The problem (\ref{eq:pro}) is equivalent to
\begin{equation}\label{eq:pro1}
\begin{array}{ll}
\mathrm{minimize} & \sum_{jk}|\boldsymbol{b_{jk}}|,\\
\mathrm{subject\ to} & \sum_{jk}\boldsymbol{\omega_{jk}}\cdot\boldsymbol{b_{jk}}=c.
\end{array}
\end{equation}
This problem can be solved as
\begin{eqnarray}\label{eq:proof}
\sum_{jk}|\boldsymbol{b_{jk}}|&=&\sum_{jk}|\boldsymbol{b_{jk}}|\cdot|\boldsymbol{\omega_{jk}}|\nonumber\\
&\geq&\sum_{jk}|\boldsymbol{b_{jk}}\cdot\boldsymbol{\omega_{jk}}|\nonumber\\
&\geq&\bigg|\sum_{jk}\boldsymbol{b_{jk}}\cdot\boldsymbol{\omega_{jk}}\bigg|=|c|,
\end{eqnarray}
where the first equation is from $|\boldsymbol{\omega_{jk}}|=1$. The equations in the second and third can hold simultaneously when each $\boldsymbol{b_{jk}}$ is in the same direction as $\boldsymbol{\omega_{jk}}$ or in the opposite direction of $\boldsymbol{\omega_{jk}}$. This leads to the following theorem.
\begin{theorem}
For a given normalized coherence witness $W$ and an unknown state $\rho$, we have
\begin{equation}
C_{l_1}(\rho)\geq\big|\langle W\rangle\big|,
\end{equation}
where $\langle W\rangle:=\tr(W\rho)$ is the measurement mean value of $W$ in $\rho$.
\end{theorem}
This theorem provides a quantitative connection between the measurement result of coherence witness and the amount of coherence contained in $\rho$. Namely, if we measure a coherence witness and obtain the mean value $c$, we know that the coherence $l_1$-norm of coherence contained in $\rho$ is at least $|c|$.

\section{Optimal coherence witness and coherence game}
In the previous section, we prove that for an \emph{unknown} state $\rho$, a large measurement mean value of a normalized coherence witness provides an evidence that the coherence contained in $\rho$ is high. Consequently, we consider a related but different problem: for a \emph{known} state $\rho$, whether we can measure its coherence using only one witness. In order to solve this problem, we define the witnessed coherence as
\begin{equation}\label{eq:wit_coh}
C_W(\rho):=\max_{W\in\wit}\tr(W\rho).
\end{equation}
The normalized witness that reach the maximum of Eq. (\ref{eq:wit_coh}) is called the optimal witness of $\rho$, and labeled as $W_{opt}(\rho)$. From Eq. (\ref{eq:proof}), one can easily check that each unit vector $\boldsymbol{w_{jk}}$ in the optimal witness is along the same direction as $\boldsymbol{b_{jk}}$ in $\rho$. If $\boldsymbol{b_{jk}}=0$ for some $j,k$, the corresponding $\boldsymbol{w_{jk}}$ can be arbitrary. Hence for states with some zero off-diagonal elements, the optimal witness is not unique. Because the optimal witness only depends on the directions of the vectors $\boldsymbol{b_{jk}}$, it can be derived without knowing the exact form of $\rho$. In other words, once we know the directions of the vectors $\boldsymbol{b_{jk}}$, the exact form of $W_{opt}(\rho)$ is known and we can use it to measure $C_W(\rho)$.

Following the similar lines as in Eq. (\ref{eq:proof}), one can easily prove that the witnessed coherence of a state coincides with the $l_1$-norm of coherence
\begin{equation}\label{eq:cw_cl1}
C_W(\rho)=C_{l_1}(\rho).
\end{equation}
This equality guarantees the witnessed coherence embody all of the properties of $C_{l_1}$ \cite{Baumgratz2014Quantifying}, such as (C1) $C_W(\rho)=0$ iff $\rho\in\incoh$, (C2) $C_W(\rho)\geq\sum_np_nC_W(\rho_n)$ where $p_n=\tr(K_n\rho K_n^\dagger)$, $\rho_n=K_n\rho K_n^\dagger/p_n$, and $\{K_n\}$ are incoherent Kraus operators of an arbitrary incoherent operation, and (C3) $\sum_np_nC_W(\rho_n)\geq C_W(\sum_np_n\rho_n)$.

Eq. (\ref{eq:cw_cl1}) demonstrates an operational meaning of the $l_1$-norm of coherence, and provides a way to detect the amount of quantum coherence by measuring only one observable if we have some previous knowledge of the state. However, because measuring the witness is \emph{not} an incoherent operation and very expensive in the resource theory of quantum coherence, it is not feasible to measure the optimal witness for each state. A natural solution to this confliction is to fix the witness $W$ and apply an incoherent operation $\Lambda^I$ (which is free) to the \emph{known} state $\rho$ before the measurement. The mean value is then $\tr(W\Lambda^I(\rho))$, which is intuitively no larger than $C_W(\rho)$. The purpose of applying $\Lambda^I$ is to maximize the mean value such that it can approach $C_W(\rho)$.

Now we are facing a problem whether $\max_{\Lambda^I}\tr(W\Lambda^I(\rho))$ can reach $C_W(\rho)$ for a fixed $W$ and all choices of $\rho$. The answer is stated in the following theorem.
\begin{theorem}\label{th:3}
For $\rho\in\mathcal D(\mathcal H_d)$, $W\in\wit$, and $\Lambda^I\in\io$,
\begin{equation}\label{eq:opt_U}
\max_{\Lambda^I}\tr(W\Lambda^I(\rho))\leq C_W(\rho).
\end{equation}
The equation holds only when the orientations $\{\theta_{ij}\}$ and $\{\theta'_{ij}\}$ of $W$ and $W_{opt}(\rho)$ satisfy that there exist a one-to-one function $f(i)$ from the index set of basis such that
\begin{equation}\label{eq:relation}
\theta_{f(i)f(j)}+\theta_{f(j)f(k)}-\theta_{f(i)f(k)}=\theta'_{ij}+\theta'_{jk}-\theta'_{ik},
\end{equation}
$\forall i,j,k$ satisfying $\boldsymbol{b_{ij}},\ \boldsymbol{b_{jk}},$ and $\boldsymbol{b_{ik}}$ in $\rho$ are nonzero.
\begin{proof}
For all $\Lambda^I\in\io$ we have
\begin{eqnarray}
\tr(W\Lambda^I(\rho))\leq C_W(\Lambda^I(\rho))\leq C_W(\rho).
\end{eqnarray}
The first equality holds when $W$ is the optimal witness of $\Lambda^I(\rho)$, and the second one holds when the coherence of $\rho$ is not changed by $\Lambda^I$. Expressed in the langue of Bloch space, the equality in Eq. (\ref{eq:opt_U}) holds when $\Lambda^I$ reorder the basis according to $|j\rangle\rightarrow|f(j)\rangle$ and rotate each vector $\boldsymbol{b_{jk}}$ of $\rho$ to the direction of $\boldsymbol{w_{f(j)f(k)}}$ of $W$, so we have
\begin{equation}
\boldsymbol{w_{f(j)f(k)}}\cdot\rot_{jk}(\boldsymbol{b_{jk}})=|\boldsymbol{b_{jk}}|=\boldsymbol{w'_{jk}}\cdot\boldsymbol{b_{jk}},
\end{equation}
where $\boldsymbol{w'_{jk}}$ are the vector of $W_{opt}(\rho)$. Notice $\boldsymbol{w_{f(j)f(k)}}\cdot\rot_{jk}(\boldsymbol{b_{jk}})=\rot_{jk}^{-1}(\boldsymbol{w_{f(j)f(k)}})\cdot\boldsymbol{b_{jk}}$, and then the equality in Eq. (\ref{eq:opt_U}) holds when the rotations $\rot_{jk}$ and the one-to-one index function $f(j)$ exist such that
\begin{eqnarray}\label{eq:condition}
\boldsymbol{w_{f(j)f(k)}}=\rot_{jk}(\boldsymbol{w'_{jk}}),\ \forall\ j,k\ \mathrm{s.t.}\ \boldsymbol{b_{jk}}\neq0.
\end{eqnarray}
Unfortunately, the rotations $\rot_{jk}$ are not independent from each other. Certain relations of $\boldsymbol{w_{f(j)f(k)}}$ and $\boldsymbol{w'_{jk}}$ should be matched if all of the equations hold. In order to derive the relations, we notice that the rotations and the reorder of basis are equivalent to an incoherent unitary operator $U_I=\sum_je^{i\lambda_j}|f(j)\rangle\langle j|$. Eq. (\ref{eq:condition}) is then equivalent to
\begin{equation}
W=U_IW_{opt}U_I^\dagger.
\end{equation}
By expanding the two sides on the basis $\{\sigma_s^{jk},\sigma_a^{jk}\}$ and comparing the coefficients, we finally arrive at Eq. (\ref{eq:relation}).
\end{proof}
\end{theorem}

If there is only one or two nonzero $\boldsymbol{b_{jk}}$ in $\rho$, the equality in Eq. (\ref{eq:opt_U}) holds for any normalized witness. For example, when $\rho$ is a single qubit state $\rho=\frac12(\iden+r\cos\theta'\sigma_x+r\sin\theta'\sigma_y+r_z\sigma_z),\ (r>0)$ and the witness is fixed as $W=\cos\theta\sigma_x+\sin\theta\sigma_y$, the incoherent unitary operator is chosen as $U_I=|0\rangle\langle0|+e^{i(\theta-\theta')}|1\rangle\langle1|$. Then we have $\tr(WU_I\rho U_I^\dagger)=r=C_W(\rho)$.

Next, we provide an example to show that the left-hand-side of Eq. (\ref{eq:opt_U}) can be strictly smaller than the right-hand-side. Here we choose the state to be measured as a pure state
\begin{equation}
\label{eq:example_rho}|\psi\rangle=\frac{\cos\theta}{\sqrt2}(|0\rangle+|1\rangle)+\sin\theta|2\rangle
\end{equation}
with $\theta\in[0,\frac{\pi}{2}]$, and the fixed witness as
\begin{equation}
\label{eq:example_W}W=\sigma_s^{01}+\sigma_s^{12}-\sigma_a^{02}.
\end{equation}
The witnessed coherence of $\rho$ is calculated using Eqs. (\ref{eq:c1}) and (\ref{eq:cw_cl1})
\begin{equation}
C_W(\rho)=\cos^2\theta+\sqrt2\sin2\theta.
\end{equation}
Our purpose here is to apply an incoherent unitary operator on $\rho$ such that the measured mean value of $W$ is as large as possible. From the symmetry in the witness and the state, the general form of the incoherent unitary operator can be written as
\begin{equation}
U_I=|0\rangle\langle0|+e^{i\lambda_1}|1\rangle\langle1|+e^{i\lambda_2}|2\rangle\langle2|
\end{equation}
with $\lambda_1,\lambda_2\in[0,2\pi)$. The mean value of the given witness $W$ after the action of unitary operator is
\begin{equation}
\begin{aligned}
\tr(WU_I\rho U_I^\dagger)=&\cos^2\theta\cos\lambda_1-\sqrt2\cos\theta\sin\theta\sin\lambda_2\\
&\sqrt2\cos\theta\sin\theta\cos(\lambda_2-\lambda_1).
\end{aligned}
\end{equation}
We maximize the mean value over $\lambda_1$ and $\lambda_2$ using numerical method, and plot the maximized mean value, as well as the witnessed coherence, in Fig. 1.

\begin{figure}\label{fig:1}
\begin{center}
\small
\centering
\includegraphics[width=9 cm]{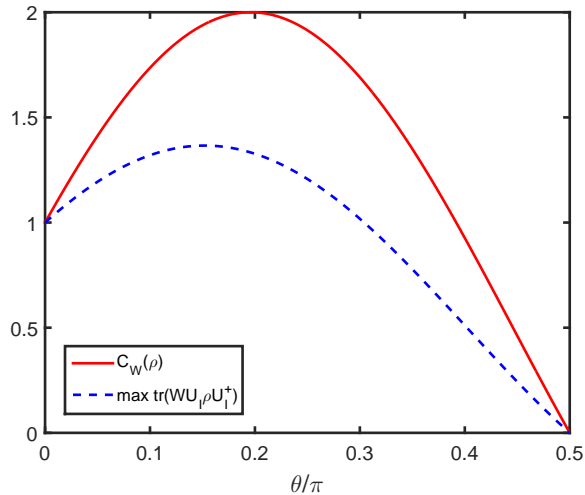}
\caption{The lhs and rhs of Eq. (\ref{eq:opt_U}) as a function of $p$. The fixed witness and the state are in the form of Eqs. (\ref{eq:example_W}) and (\ref{eq:example_rho}).}	
\end{center}
\end{figure}

A direct application of Theorem \ref{th:3} is the following quantum game. In this game, the referee controls a detector which measures the observable $W$ as in Eq. (\ref{w_d_gen}). The player has a known state $\rho$ and he can implement any incoherent unitary operator $U_I$ to it. After he optimizes his state, he put it in the referee's detector. If the measurement result is a positive number $a$, the player can earn $a$ dollars; if a negative number $-b$, he would lose $b$ dollars. The average payoff function is then
\begin{equation}
\mathcal P(\rho)=\tr(W U_I\rho U_I^\dagger).
\end{equation}
From Theorem \ref{th:3}, the player can earn some money on average as long as he has some initial coherence, and the maximal average gain is upper-bounded by the coherence of his initial state. If the player's state is a single-qubit one, the maximal average gain equals exactly to the initial coherence. When higher dimension is considered, the average payoff cannot reach the initial coherence if the orientations of $\rho$ and $W$ do not match. It means that, some amount of coherence in quantum systems of dimension higher than two can not be activated in this game.

\section{Conclusion}
The stringent coherence witness, and several related optimization problems have been investigated. The witness is stringent in the way that all of the incoherent states should satisfy the equation $\tr(W\rho_I)=0$, instead of the inequality $\tr(\tilde W\rho_I)\geq0$ for the traditional witnesses. We prove that such witness exist for arbitrary finite-dimension coherent states.

When the coherence witness is normalized, its mean value reveals the amount of coherence in the measured state. For an unknown state and a fixed witness, the modulus of the mean value provides a tight lower bound of the $l_1$-norm of coherence contained in the state. For a given state $\rho$, the optimal witness is defined as the normalized witness whose mean value in $\rho$ reaches the maximum. Once the orientations of the vectors $\boldsymbol{b_{jk}}$ in $\rho$ is known, the optimal witness is fixed. Hence one can directly measure $W_{opt}$ and the mean value is just the $l_1$-norm of coherence in $\rho$.

Another important problem goes as follows. Using a fixed $W$, can we measure the coherence of a state if its $\boldsymbol{b_{jk}}$ are known and incoherent operations are allowed? In the qubit case, the answer is yes. But in the high-dimension cases, the coherence can be measured only when certain relations (as in Eq. (\ref{eq:relation})) between $\boldsymbol{w_{jk}}$ in $W$ and $\boldsymbol{b_{jk}}$ is satisfied. We study an example to show the gap between the maximum mean value of $W$ and the witnessed coherence of the state. A coherence game is also designed to show this effect.

\begin{acknowledgments}
XH thanks the helpful discussions with D. L. Zhou. This work was supported by NSFC under Grant No. 11504205.
\end{acknowledgments}

%In this study, we define an improved coherent witness to argue the coherence contained in an unknown state without a full tomographic knowledge about the state. On the basis of expectation values of coherent witnesses, we calculate the minimal coherence contained in the state and get the number of witnesses needed to judge whether the state is coherent. In this way, we measure the expectation values of $W$ frequently in oder to get the coherence more accurately. Besides, we have considered incoherent unitary operations which have great implications.

%\newpage %Just because of unusual number of tables stacked at end
%\bibliographystyle{unsrt}
\bibliography{coherence}

\begin{thebibliography}{27}
\expandafter\ifx\csname natexlab\endcsname\relax\def\natexlab#1{#1}\fi
\expandafter\ifx\csname bibnamefont\endcsname\relax
  \def\bibnamefont#1{#1}\fi
\expandafter\ifx\csname bibfnamefont\endcsname\relax
  \def\bibfnamefont#1{#1}\fi
\expandafter\ifx\csname citenamefont\endcsname\relax
  \def\citenamefont#1{#1}\fi
\expandafter\ifx\csname url\endcsname\relax
  \def\url#1{\texttt{#1}}\fi
\expandafter\ifx\csname urlprefix\endcsname\relax\def\urlprefix{URL }\fi
\providecommand{\bibinfo}[2]{#2}
\providecommand{\eprint}[2][]{\url{#2}}

\bibitem[{\citenamefont{Lewenstein et~al.}(2001)\citenamefont{Lewenstein,
  Kraus, Horodecki, and Cirac}}]{Lewenstein2000Characterization}
\bibinfo{author}{\bibfnamefont{M.}~\bibnamefont{Lewenstein}},
  \bibinfo{author}{\bibfnamefont{B.}~\bibnamefont{Kraus}},
  \bibinfo{author}{\bibfnamefont{P.}~\bibnamefont{Horodecki}},
  \bibnamefont{and} \bibinfo{author}{\bibfnamefont{J.~I.} \bibnamefont{Cirac}},
  \bibinfo{journal}{Phys. Rev. A} \textbf{\bibinfo{volume}{63}},
  \bibinfo{pages}{044304} (\bibinfo{year}{2001}).

\bibitem[{\citenamefont{Eisert et~al.}(2007)\citenamefont{Eisert, Brand\~ao,
  and Audenaert}}]{Eisert2006Quantitative}
\bibinfo{author}{\bibfnamefont{J.}~\bibnamefont{Eisert}},
  \bibinfo{author}{\bibfnamefont{F.~G. S.~L.} \bibnamefont{Brand\~ao}},
  \bibnamefont{and} \bibinfo{author}{\bibfnamefont{K.~M.~R.}
  \bibnamefont{Audenaert}}, \bibinfo{journal}{New Journal of Physics}
  \textbf{\bibinfo{volume}{9}}, \bibinfo{pages}{46} (\bibinfo{year}{2007}).

\bibitem[{\citenamefont{Brand\~ao}(2005)}]{Brand2005Quantifying}
\bibinfo{author}{\bibfnamefont{F.~G. S.~L.} \bibnamefont{Brand\~ao}},
  \bibinfo{journal}{Phys. Rev. A} \textbf{\bibinfo{volume}{72}},
  \bibinfo{pages}{022310} (\bibinfo{year}{2005}).

\bibitem[{\citenamefont{Buscemi}(2012)}]{PhysRevLett.108.200401}
\bibinfo{author}{\bibfnamefont{F.}~\bibnamefont{Buscemi}},
  \bibinfo{journal}{Phys. Rev. Lett.} \textbf{\bibinfo{volume}{108}},
  \bibinfo{pages}{200401} (\bibinfo{year}{2012}).

\bibitem[{\citenamefont{Branciard et~al.}(2013)\citenamefont{Branciard, Rosset,
  Liang, and Gisin}}]{PhysRevLett.110.060405}
\bibinfo{author}{\bibfnamefont{C.}~\bibnamefont{Branciard}},
  \bibinfo{author}{\bibfnamefont{D.}~\bibnamefont{Rosset}},
  \bibinfo{author}{\bibfnamefont{Y.-C.} \bibnamefont{Liang}}, \bibnamefont{and}
  \bibinfo{author}{\bibfnamefont{N.}~\bibnamefont{Gisin}},
  \bibinfo{journal}{Phys. Rev. Lett.} \textbf{\bibinfo{volume}{110}},
  \bibinfo{pages}{060405} (\bibinfo{year}{2013}).

\bibitem[{\citenamefont{Cavalcanti et~al.}(2013)\citenamefont{Cavalcanti, Hall,
  and Wiseman}}]{PhysRevA.87.032306}
\bibinfo{author}{\bibfnamefont{E.~G.} \bibnamefont{Cavalcanti}},
  \bibinfo{author}{\bibfnamefont{M.~J.~W.} \bibnamefont{Hall}},
  \bibnamefont{and} \bibinfo{author}{\bibfnamefont{H.~M.}
  \bibnamefont{Wiseman}}, \bibinfo{journal}{Phys. Rev. A}
  \textbf{\bibinfo{volume}{87}}, \bibinfo{pages}{032306}
  (\bibinfo{year}{2013}).

\bibitem[{\citenamefont{Chen et~al.}(2017)\citenamefont{Chen, Hu, and
  Zhou}}]{PhysRevA.95.052326}
\bibinfo{author}{\bibfnamefont{X.}~\bibnamefont{Chen}},
  \bibinfo{author}{\bibfnamefont{X.}~\bibnamefont{Hu}}, \bibnamefont{and}
  \bibinfo{author}{\bibfnamefont{D.~L.} \bibnamefont{Zhou}},
  \bibinfo{journal}{Phys. Rev. A} \textbf{\bibinfo{volume}{95}},
  \bibinfo{pages}{052326} (\bibinfo{year}{2017}).

\bibitem[{\citenamefont{Lewenstein et~al.}(2000)\citenamefont{Lewenstein,
  Kraus, Cirac, and Horodecki}}]{Lewenstein2000Optimization}
\bibinfo{author}{\bibfnamefont{M.}~\bibnamefont{Lewenstein}},
  \bibinfo{author}{\bibfnamefont{B.}~\bibnamefont{Kraus}},
  \bibinfo{author}{\bibfnamefont{J.~I.} \bibnamefont{Cirac}}, \bibnamefont{and}
  \bibinfo{author}{\bibfnamefont{P.}~\bibnamefont{Horodecki}},
  \bibinfo{journal}{Phys. Rev. A} \textbf{\bibinfo{volume}{62}},
  \bibinfo{pages}{052310} (\bibinfo{year}{2000}).

\bibitem[{\citenamefont{Shahandeh et~al.}(2017)\citenamefont{Shahandeh,
  Ringbauer, Loredo, and Ralph}}]{PhysRevLett.118.110502}
\bibinfo{author}{\bibfnamefont{F.}~\bibnamefont{Shahandeh}},
  \bibinfo{author}{\bibfnamefont{M.}~\bibnamefont{Ringbauer}},
  \bibinfo{author}{\bibfnamefont{J.~C.} \bibnamefont{Loredo}},
  \bibnamefont{and} \bibinfo{author}{\bibfnamefont{T.~C.} \bibnamefont{Ralph}},
  \bibinfo{journal}{Phys. Rev. Lett.} \textbf{\bibinfo{volume}{118}},
  \bibinfo{pages}{110502} (\bibinfo{year}{2017}).

\bibitem[{\citenamefont{Streltsov et~al.}()\citenamefont{Streltsov, Adesso, and
  Plenio}}]{arXiv:1609.02439}
\bibinfo{author}{\bibfnamefont{A.}~\bibnamefont{Streltsov}},
  \bibinfo{author}{\bibfnamefont{G.}~\bibnamefont{Adesso}}, \bibnamefont{and}
  \bibinfo{author}{\bibfnamefont{M.~B.} \bibnamefont{Plenio}},
  \bibinfo{note}{arXiv:1609.02439}.

\bibitem[{\citenamefont{Hu et~al.}()\citenamefont{Hu, Hu, Peng, Zhang, and
  Fan}}]{arXiv:1703.01852}
\bibinfo{author}{\bibfnamefont{M.-L.} \bibnamefont{Hu}},
  \bibinfo{author}{\bibfnamefont{X.}~\bibnamefont{Hu}},
  \bibinfo{author}{\bibfnamefont{Y.}~\bibnamefont{Peng}},
  \bibinfo{author}{\bibfnamefont{Y.-R.} \bibnamefont{Zhang}}, \bibnamefont{and}
  \bibinfo{author}{\bibfnamefont{H.}~\bibnamefont{Fan}},
  \bibinfo{note}{arXiv:1703.01852}.

\bibitem[{\citenamefont{Streltsov et~al.}(2015)\citenamefont{Streltsov, Singh,
  Dhar, Bera, and Adesso}}]{PhysRevLett.115.020403}
\bibinfo{author}{\bibfnamefont{A.}~\bibnamefont{Streltsov}},
  \bibinfo{author}{\bibfnamefont{U.}~\bibnamefont{Singh}},
  \bibinfo{author}{\bibfnamefont{H.~S.} \bibnamefont{Dhar}},
  \bibinfo{author}{\bibfnamefont{M.~N.} \bibnamefont{Bera}}, \bibnamefont{and}
  \bibinfo{author}{\bibfnamefont{G.}~\bibnamefont{Adesso}},
  \bibinfo{journal}{Phys. Rev. Lett.} \textbf{\bibinfo{volume}{115}},
  \bibinfo{pages}{020403} (\bibinfo{year}{2015}).

\bibitem[{\citenamefont{Hu and Fan}(2016)}]{Hu16}
\bibinfo{author}{\bibfnamefont{X.}~\bibnamefont{Hu}} \bibnamefont{and}
  \bibinfo{author}{\bibfnamefont{H.}~\bibnamefont{Fan}}, \bibinfo{journal}{Sci.
  Rep.} \textbf{\bibinfo{volume}{6}}, \bibinfo{pages}{34380}
  (\bibinfo{year}{2016}).

\bibitem[{\citenamefont{Ma et~al.}(2016)\citenamefont{Ma, Yadin, Girolami,
  Vedral, and Gu}}]{PhysRevLett.116.160407}
\bibinfo{author}{\bibfnamefont{J.}~\bibnamefont{Ma}},
  \bibinfo{author}{\bibfnamefont{B.}~\bibnamefont{Yadin}},
  \bibinfo{author}{\bibfnamefont{D.}~\bibnamefont{Girolami}},
  \bibinfo{author}{\bibfnamefont{V.}~\bibnamefont{Vedral}}, \bibnamefont{and}
  \bibinfo{author}{\bibfnamefont{M.}~\bibnamefont{Gu}}, \bibinfo{journal}{Phys.
  Rev. Lett.} \textbf{\bibinfo{volume}{116}}, \bibinfo{pages}{160407}
  (\bibinfo{year}{2016}).

\bibitem[{\citenamefont{Chitambar and Hsieh}(2016)}]{PhysRevLett.117.020402}
\bibinfo{author}{\bibfnamefont{E.}~\bibnamefont{Chitambar}} \bibnamefont{and}
  \bibinfo{author}{\bibfnamefont{M.-H.} \bibnamefont{Hsieh}},
  \bibinfo{journal}{Phys. Rev. Lett.} \textbf{\bibinfo{volume}{117}},
  \bibinfo{pages}{020402} (\bibinfo{year}{2016}).

\bibitem[{\citenamefont{Baumgratz et~al.}(2014)\citenamefont{Baumgratz, Cramer,
  and Plenio}}]{Baumgratz2014Quantifying}
\bibinfo{author}{\bibfnamefont{T.}~\bibnamefont{Baumgratz}},
  \bibinfo{author}{\bibfnamefont{M.}~\bibnamefont{Cramer}}, \bibnamefont{and}
  \bibinfo{author}{\bibfnamefont{M.~B.} \bibnamefont{Plenio}},
  \bibinfo{journal}{Phys. Rev. Lett.} \textbf{\bibinfo{volume}{113}},
  \bibinfo{pages}{140401} (\bibinfo{year}{2014}).

\bibitem[{\citenamefont{Winter and Yang}(2016)}]{arXiv:1506.07975}
\bibinfo{author}{\bibfnamefont{A.}~\bibnamefont{Winter}} \bibnamefont{and}
  \bibinfo{author}{\bibfnamefont{D.}~\bibnamefont{Yang}},
  \bibinfo{journal}{Phys. Rev. Lett.} \textbf{\bibinfo{volume}{116}},
  \bibinfo{pages}{120404} (\bibinfo{year}{2016}).

\bibitem[{\citenamefont{Yuan et~al.}(2015)\citenamefont{Yuan, Zhou, Cao, and
  Ma}}]{PhysRevA.92.022124}
\bibinfo{author}{\bibfnamefont{X.}~\bibnamefont{Yuan}},
  \bibinfo{author}{\bibfnamefont{H.}~\bibnamefont{Zhou}},
  \bibinfo{author}{\bibfnamefont{Z.}~\bibnamefont{Cao}}, \bibnamefont{and}
  \bibinfo{author}{\bibfnamefont{X.}~\bibnamefont{Ma}}, \bibinfo{journal}{Phys.
  Rev. A} \textbf{\bibinfo{volume}{92}}, \bibinfo{pages}{022124}
  (\bibinfo{year}{2015}).

\bibitem[{\citenamefont{Napoli et~al.}(2016)\citenamefont{Napoli, Bromley,
  Cianciaruso, Piani, Johnston, and Adesso}}]{Napoli2016Robustness}
\bibinfo{author}{\bibfnamefont{C.}~\bibnamefont{Napoli}},
  \bibinfo{author}{\bibfnamefont{T.~R.} \bibnamefont{Bromley}},
  \bibinfo{author}{\bibfnamefont{M.}~\bibnamefont{Cianciaruso}},
  \bibinfo{author}{\bibfnamefont{M.}~\bibnamefont{Piani}},
  \bibinfo{author}{\bibfnamefont{N.}~\bibnamefont{Johnston}}, \bibnamefont{and}
  \bibinfo{author}{\bibfnamefont{G.}~\bibnamefont{Adesso}},
  \bibinfo{journal}{Phys. Rev. Lett.} \textbf{\bibinfo{volume}{116}},
  \bibinfo{pages}{150502} (\bibinfo{year}{2016}).

\bibitem[{\citenamefont{Yu et~al.}(2016)\citenamefont{Yu, Zhang, Xu, and
  Tong}}]{PhysRevA.94.060302}
\bibinfo{author}{\bibfnamefont{X.-D.} \bibnamefont{Yu}},
  \bibinfo{author}{\bibfnamefont{D.-J.} \bibnamefont{Zhang}},
  \bibinfo{author}{\bibfnamefont{G.~F.} \bibnamefont{Xu}}, \bibnamefont{and}
  \bibinfo{author}{\bibfnamefont{D.~M.} \bibnamefont{Tong}},
  \bibinfo{journal}{Phys. Rev. A} \textbf{\bibinfo{volume}{94}},
  \bibinfo{pages}{060302} (\bibinfo{year}{2016}).

\bibitem[{\citenamefont{Liu et~al.}(2017)\citenamefont{Liu, Hu, and
  Lloyd}}]{PhysRevLett.118.060502}
\bibinfo{author}{\bibfnamefont{Z.-W.} \bibnamefont{Liu}},
  \bibinfo{author}{\bibfnamefont{X.}~\bibnamefont{Hu}}, \bibnamefont{and}
  \bibinfo{author}{\bibfnamefont{S.}~\bibnamefont{Lloyd}},
  \bibinfo{journal}{Phys. Rev. Lett.} \textbf{\bibinfo{volume}{118}},
  \bibinfo{pages}{060502} (\bibinfo{year}{2017}).

\bibitem[{\citenamefont{Rana et~al.}(2016)\citenamefont{Rana, Parashar, and
  Lewenstein}}]{PhysRevA.93.012110}
\bibinfo{author}{\bibfnamefont{S.}~\bibnamefont{Rana}},
  \bibinfo{author}{\bibfnamefont{P.}~\bibnamefont{Parashar}}, \bibnamefont{and}
  \bibinfo{author}{\bibfnamefont{M.}~\bibnamefont{Lewenstein}},
  \bibinfo{journal}{Phys. Rev. A} \textbf{\bibinfo{volume}{93}},
  \bibinfo{pages}{012110} (\bibinfo{year}{2016}).

\bibitem[{\citenamefont{Hu}(2016)}]{PhysRevA.94.012326}
\bibinfo{author}{\bibfnamefont{X.}~\bibnamefont{Hu}}, \bibinfo{journal}{Phys.
  Rev. A} \textbf{\bibinfo{volume}{94}}, \bibinfo{pages}{012326}
  (\bibinfo{year}{2016}).

\bibitem[{\citenamefont{Piani et~al.}(2016)\citenamefont{Piani, Cianciaruso,
  Bromley, Napoli, Johnston, and Adesso}}]{PhysRevA.93.042107}
\bibinfo{author}{\bibfnamefont{M.}~\bibnamefont{Piani}},
  \bibinfo{author}{\bibfnamefont{M.}~\bibnamefont{Cianciaruso}},
  \bibinfo{author}{\bibfnamefont{T.~R.} \bibnamefont{Bromley}},
  \bibinfo{author}{\bibfnamefont{C.}~\bibnamefont{Napoli}},
  \bibinfo{author}{\bibfnamefont{N.}~\bibnamefont{Johnston}}, \bibnamefont{and}
  \bibinfo{author}{\bibfnamefont{G.}~\bibnamefont{Adesso}},
  \bibinfo{journal}{Phys. Rev. A} \textbf{\bibinfo{volume}{93}},
  \bibinfo{pages}{042107} (\bibinfo{year}{2016}).

\bibitem[{\citenamefont{Wang et~al.}(2017)\citenamefont{Wang, Tang, Wei, Yu,
  Ke, Xu, Li, and Guo}}]{PhysRevLett.118.020403}
\bibinfo{author}{\bibfnamefont{Y.-T.} \bibnamefont{Wang}},
  \bibinfo{author}{\bibfnamefont{J.-S.} \bibnamefont{Tang}},
  \bibinfo{author}{\bibfnamefont{Z.-Y.} \bibnamefont{Wei}},
  \bibinfo{author}{\bibfnamefont{S.}~\bibnamefont{Yu}},
  \bibinfo{author}{\bibfnamefont{Z.-J.} \bibnamefont{Ke}},
  \bibinfo{author}{\bibfnamefont{X.-Y.} \bibnamefont{Xu}},
  \bibinfo{author}{\bibfnamefont{C.-F.} \bibnamefont{Li}}, \bibnamefont{and}
  \bibinfo{author}{\bibfnamefont{G.-C.} \bibnamefont{Guo}},
  \bibinfo{journal}{Phys. Rev. Lett.} \textbf{\bibinfo{volume}{118}},
  \bibinfo{pages}{020403} (\bibinfo{year}{2017}).

\bibitem[{\citenamefont{Horodecki et~al.}(2000)\citenamefont{Horodecki,
  Horodecki, and Horodecki}}]{Horodecki2000Separability}
\bibinfo{author}{\bibfnamefont{M.}~\bibnamefont{Horodecki}},
  \bibinfo{author}{\bibfnamefont{P.}~\bibnamefont{Horodecki}},
  \bibnamefont{and}
  \bibinfo{author}{\bibfnamefont{R.}~\bibnamefont{Horodecki}},
  \bibinfo{journal}{Physics Letters A} \textbf{\bibinfo{volume}{283}},
  \bibinfo{pages}{1} (\bibinfo{year}{2000}).

\bibitem[{\citenamefont{Bertlmann and Krammer}(2008)}]{1751-8121-41-23-235303}
\bibinfo{author}{\bibfnamefont{R.~A.} \bibnamefont{Bertlmann}}
  \bibnamefont{and} \bibinfo{author}{\bibfnamefont{P.}~\bibnamefont{Krammer}},
  \bibinfo{journal}{Journal of Physics A: Mathematical and Theoretical}
  \textbf{\bibinfo{volume}{41}}, \bibinfo{pages}{235303}
  (\bibinfo{year}{2008}).

\end{thebibliography}
\end{document}